\documentclass[10pt]{article}

\newcommand{\sect}[1]{\setcounter{equation}{0}\section{#1}}
\renewcommand{\theequation}{\arabic{section}.\arabic{equation}}

\def\be{\begin{equation}}
\def\ee{\end{equation}}
\def\bea{\begin{eqnarray}}
\def\eea{\end{eqnarray}}
\def\nn{\nonumber \\ [.2cm]}
\def\vsp#1{\vspace{#1}}
\def\hsp#1{\hspace{#1}}

\def\part{\partial}
\def\tfrac#1#2{{\textstyle{\frac{#1}{#2}}}}

\def\Tr{\mbox{Tr}}

\def\incl{{\mbox{\footnotesize i}}}
\def\ii{{\incl_X \incl_X}}

\def\cF{{\cal F}}

\def\mn{{\mu\nu}}

\def\makeatletter{\catcode`\@=11}
\makeatletter
\def\mathbox#1{\hbox{$\m@th#1$}}%
\def\math@ccstyles#1#2#3#4#5#6#7{{\leavevmode
       \setbox0\mathbox{#6#7}%
       \setbox2\mathbox{#4#5}%
       \dimen@ #3%
       \baselineskip\z@\lineskiplimit#1\lineskip\z@
       \vbox{\ialign{##\crcr
              \hfil \kern #2\box2 \hfil\crcr
              \noalign{\kern\dimen@}%
              \hfil\box0\hfil\crcr}}}}
\def\mathaccstyles{\math@ccstyles\maxdimen}
\def\maththroughstyles{\math@ccstyles{-\maxdimen}}
\def\unity%
  {\maththroughstyles{.45\ht0}\z@\displaystyle
{\mathchar"006C}\displaystyle 1}

\begin{document}

\rightline{KUL-TF/05-26}
\rightline{hep-th/0511191}
\rightline{November 2005}
\vspace{1truecm}

\centerline{\LARGE \bf Comparing two definitions for gauge variations  }
\vspace{.5truecm}
\centerline{\LARGE \bf of multiple D-branes}
\vspace{1.3truecm}

\centerline{
    {\large \bf Joke Adam}\footnote{E-mail address: 
                                  {\tt joke.adam@fys.kuleuven.ac.be}},
                                                            }
\vspace{.4cm}
\centerline{{\it Instituut voor Theoretische Fysica, K.U. Leuven,}}
\centerline{{\it Celestijnenlaan 200D,  B-3001 Leuven, Belgium}}

\vspace{2truecm}


\centerline{\bf ABSTRACT}
\vspace{.5truecm}
\noindent
We compare two definitions of gauge variations in the case of non-Abelian actions for multiple D-branes. Equivalence is proven for the R-R variations, which shows that the action is invariant also under the easier, naive variation. For the NS-NS variations however, the two definitions are not equivalent, leaving the naive definition as the only valid one.
\newpage
\sect{Introduction}\label{sectintro}
Parallel D-branes can behave collectively rather than independently, providing strange, non-Abelian physics. The phenomenon occurs when the distance between the branes becomes of order of the string length. Then the strings stretching between them \cite{Witten} have massless states, in addition to the already present massless states of the strings going from a brane to itself. In the worldvolume description this corresponds to the $N$ $U(1)$ groups, one for each brane, filling out to one $U(N)$ symmetry. So the $N$ $U(1)$ Born-Infeld vectors become arranged into one $U(N)$ Yang-Mills vector $V$ while the transverse scalars become non-Abelian $U(N)$ matrices $X^i$. These matrix coordinates still contain the information about the coordinates of the distinct branes: the $I$-th eigenvalue of $X^{i}$ is the $i$-th coordinate of the $I$-th brane. In general however, $U(N)$ matrices can not be diagonalized simultaneously, such that an uncertainty exists on the coordinates.

The worldvolume action of such a multiple brane should encode the physics resulting from the non-Abelian structure. Defining a Born-Infeld action is a highly non-trivial problem, but the Chern-Simons part seems to keep a simple structure. Still, also the latter receives some important modifications. The first generalization of the Chern-Simons term to the $U(N)$ case consisted of defining the Born-Infeld field strength $F$ as a $U(N)$ field strength $F = 2\partial V + i[V,V]$ and adding a trace to the action \cite{Douglas}:
\begin{eqnarray}
S_{{\rm D}p}\ =\ T_p \int P[C] \ \Tr \{ e^\cF \}
            \ = \ T_p \int \sum_n P[C_{p-2n+1}]\  \Tr \{ \cF^n \},
\label{GHTac}
\end{eqnarray}
where $\cF$ is defined as $\cF = F + B$.

Then it was observed that the background fields should depend on the matrix coordinates via a non-Abelian Taylor expansion \cite{Douglas, GM}:
\begin{eqnarray}
C_\mn (x^a, X^i) = \sum_n \frac{1}{n!} \part_{k_1} ... \part_{k_n}
                     C_\mn (x^a, x^i){\mid}_{x^i = 0} \
                      X^{k_1} ... \ X^{k_n}.
\label{taylorexp}
\end{eqnarray}
Together with a change from ordinary world-volume derivatives to covariant derivatives $D_a = \partial_a + i[V_a, .]$ \cite{Dorn, Hull}, and the symmetrized trace prescription \cite{Tseytlin}, here denoted by $STr$, the action becomes invariant under the $U(N)$ symmetry.
The resulting action, though, does not fit with T-duality \cite{Myers}. Indeed, T-duality requires extra terms proportional to commutators between the transverse coordinates $[X,X]$. The final multiple $D$-brane Chern-Simons action looks like:
\begin{eqnarray}
S_{{\rm D}p}\ =\ T_p \int \  STr\Bigl\{ P[e^{(\incl_X \incl_X)} (C e^B) ] \  e^F \Bigr\},
\label{myersactie}
\end{eqnarray}
where $(\ii)$ stands for inclusion with the transverse scalars, $(\ii)C_p = \frac12 [X^{\rho}, X^{\sigma}]C_{\sigma \rho \mu_1 ... \mu_{p-2}}$. The same result was found for $D0$ branes using matrix theory techniques \cite{TvR1, TvR2}.
The new couplings proportional to the commutators allow the brane to interact with background fields of rank $n$ higher than the brane dimension $p+1$. Due to the non-Abelian couplings, fuzzy brane solutions \cite{Myers, TV}, which react like a dielectric to the higher rank background fields, become possible.

Gauge invariance of the multiple brane action is discussed in \cite{vR,AJGL}. In the latter it is argued that the non-Abelian pullback affected the question of gauge invariance. Indeed, naively filling in the variation $\delta C_{\mu \nu} = 2\partial_{[ \mu} \Lambda_{\nu ]}$ into the pullbacked field yields:
\begin{eqnarray}
\delta\ STr\Bigl\{ D_a X^{\mu} D_b X^{\nu} C_{\mu \nu} \Bigr\} &=& STr\Bigl\{ D_a X^{\mu} D_b X^{\nu} \partial_{[\mu}\Lambda_{\nu]} \Bigr\}.\label{arg1}
\end{eqnarray}
This is not a total derivative, which would mean that e.g. the truncation of the $D6$ brane action where all fields but $C_7$ are zero, would not be invariant under R-R gauge transformations.
Therefore R-R transformation of pullbacked fields was redefined as follows:
\begin{eqnarray}
\delta \ STr\Bigl\{ P[C_2] \Bigr\} &=& 2 \partial \ STr\Bigl\{ P[\Lambda_1] \Bigr\} = STr \Bigl\{2 DP[\Lambda_1] \Bigr\},\label{arg2}
\end{eqnarray}
which is by definition a total derivative. By analogy, full transformations are defined also for the dielectric fields\footnote{The square brackets in the summation denote the integer part.}:
\begin{eqnarray}
\delta_\Lambda STr \{ P[(\incl_X \incl_X) C_{p}] \}
        &=& \sum_{n=0}^{[(p-1)/2]} STr \Bigl\{
              \tfrac{(p-2)!}{2^n n! (p-2n-3)!} D P[(\ii)\Lambda_{p-2n-1}] P[B^n] \label{Lambdadielec} \\
      && \hsp{1cm}
            +\ \tfrac{(p-2)!}{2^{n-2} (n-1)! (p-2n-2)!} D P [\incl_X\Lambda_{p-2n-1}]
                                     P[(\incl_X B)B^{n-1}]  \nonumber
                                                             \\[.2cm]
       && \hsp{1cm}
            +\ \tfrac{(p-2)!}{2^{n-2} (n-2)! (p-2n-1)!} D P [\Lambda_{p-2n-1}]
                                     P[(\incl_X B)^2 B^{n-2}]  \nn
        && \hsp{1cm}
            +\ \tfrac{(p-2)!}{2^{n-1} (n-1)! (p-2n-1)!} D P [\Lambda_{p-2n-1}]
                                     P[(\incl_X \incl_X B) B^{n-1}] \nonumber
                            \Bigr\}.
\end{eqnarray}

These modified variations led directly to the invariance of the multiple D-brane action with respect to R-R gauge variations.
The question of invariance under NS-NS transformations is more subtle, because it is directly linked to coordinate transformations. Defining matrix coordinate transformations is still a unsolved question, though quite some progress is made by \cite{dBS, BFLvR, BKLvR}. However, using T-dualities the NS-NS variations can be fully derived without the need of coordinate transformations. T-duality performed on the Born-Infeld vector $V$ and its variation $\delta V_a = -\Sigma_{\rho}D_a X^{\rho}$ led to the discovery that the matrix coordinates are affected by a gauge variation of the form \cite{AIJ}
\begin{eqnarray}
\delta X^{\mu} &=& i\Sigma_{\rho}[X^{\rho},X^{\mu}].
\end{eqnarray}
This causes also the pullbacks, the commutators $[X,X]$ and the background fields, which depend on the matrix coordinates, to transform under $\Sigma$. For example, the variation of the NS-NS twoform $B$ is
\begin{eqnarray}
\delta P[B]&=& 2P[\partial \Sigma] - 12i P[(\ii)(B \partial \Sigma)] + 2i (\ii)B \ P[\partial \Sigma].\label{naiveB}
\end{eqnarray}
Besides the transformations due to the dependence on the coordinates, which are the two last terms, we see that the proper variation has not been changed to a modified form such as (\ref{arg2}). Instead, the NS-NS variation follows what we will call further on the naive definition: varying every factor ($X$-dependence of the fields, pullbacks, commutators, the $B$ field) and putting them just together. 
As a consequence of the different definitions, two forms which are related by S-duality, namely $C_2$ and $B$, have different gauge properties.

In this work we will compare the naive variation, which was fit for $B$, to the modified variation of the $C$'s. For the R-R variations, the naive and the modified definitions are equivalent, which will be proven in section 2 for the case of the $D6$ brane. In section 3 is taken care of the NS-NS variations, whereof a modified version does not seem to exist.  A subtlety concerning the symmetrized trace prescription is described in the appendix.

\sect{R-R transformations}
To compare the two definitions, we will look at the $D6$ brane. Its action, though being easy, has all the features of a non-Abelian Chern-Simons action:
\begin{eqnarray}
\mathcal{L}_{D6} &=& STr\Bigl\{ \sum_{r=0}^{1}\sum_{n=0}^{3}\ \frac{i^r}{r!}\frac{(-1)^{n+r}7!}{(7-2n)!2^n n!}\ P\Bigl[ (\ii)^r \mathcal{A}_{7+2r-2n} \Bigr] F^{n} \Bigr\},
\end{eqnarray}
where the background forms $\mathcal{A}_p$ are defined as
\begin{eqnarray}
\mathcal{A}_p &=& \sum_{k=0}^{[p/2]}\frac{p!}{(p-2k)!2^k k!} C_{p-2k} B^k.
\end{eqnarray}
First we look at the truncation $B=F=C_9=0$:
\begin{eqnarray}
\mathcal{L}_{D6}^{trunc1} &=& STr\Bigl\{ P[ C_7 ] \Bigr\}.
\end{eqnarray}
The modified variation is defined by being a total derivative and turns out to be a covariant derivative of the pullbacked field:
\begin{eqnarray}
\delta_{mod}\ \mathcal{L}_{D6}^{trunc1} &=& 7\partial \ STr\Bigl\{ P[ \Lambda_6 ] \Bigr\} \\
&=& STr\Bigl\{7 D \Bigl( P[ \Lambda_6 ]\Bigr) \Bigr\}. \nonumber
\end{eqnarray}
Working out this variation gives:
\begin{eqnarray}
\delta_{mod} \mathcal{L}_{D6}^{trunc1} &=& STr\Bigl\{7 P[\partial \Lambda_6] + 21i \Lambda_{\mu_1...\mu_6} [F_{[ba_1},X^{\mu_1}]D_{a_2}X^{\mu_2}... D_{a_6]}X^{\mu_6} \Bigr\}\\
&=& STr\Bigl\{7 P[\partial \Lambda_6] \Bigr\} = \delta_{naive}\ \mathcal{L}_{D6}^{trunc1}.\nonumber
\end{eqnarray}
The second term, coming from the commutator of two covariant derivatives, vanishes because we assumed $F$ to be zero. No difference is left between the naive and the modified definitions. We see here that, though (\ref{arg1}) was indeed no total derivative, invariance is still assured for the truncated action.
Allowing $F$ to be arbitrary restores the difference between the definitions. One can expect this extra term cancelling variations coming from $P[ (\ii)C_7]F$, such that
\begin{eqnarray}
\mathcal{L}_{D6}^{trunc2} &=& STr\Bigl\{ P[ C_7 ] + 21 i P[(\ii)C_7]F \Bigr\}
\end{eqnarray}
will be invariant under the naive gauge variation as well as under the modified one.
Again we will work out the modified variation and try to end up with the naive one.
\begin{eqnarray}
\delta \mathcal{L}_{D6}^{trunc2} &=& STr\Bigl\{7 D \Bigl( P[ \Lambda_6 ]\Bigr) + 21 i\cdot 5\ D \Bigl( P[(\ii)\Lambda_6]\Bigr)F  \Bigr\}\label{AJGL-naive2}\\
&=& STr\Bigl\{ 7 P[\partial \Lambda_6] + 21i \Lambda_{\mu_1...\mu_6} [F_{[ba_1}X^{\mu_1}]D_{a_2}X^{\mu_2}... D_{a_6]}X^{\mu_6} \nonumber \\
& & + 21 i \Bigl( 10i ((\ii)\Lambda)_{\mu_1...\mu_4}[F_{[ ba_1}, X^{\mu_1}]D_{a_2}X^{\mu_2}...D_{a_4}X^{\mu_4}\Bigr) F_{a_5 a_6 ]} \nonumber \\
& & + 21 i \Bigl( 5\ \Lambda_{\sigma \rho \mu_1 ... \mu_4} [D_{[b} X^{\rho}, X^{\sigma}] D_{a_1}X^{\mu_1}... D_{a_4}X^{\mu_4} \Bigr) F_{a_5 a_6 ]} \nonumber \\
& & + 21 i \Bigl( \frac12\cdot 5\ \partial_{\eta}\Lambda_{\sigma \rho \mu_1 ... \mu_4}[X^{\rho}, X^{\sigma}] D_{[b} X^{\eta} D_{a_1}X^{\mu_1}...D_{a_4}X^{\mu_4} \Bigr) F_{a_5 a_6 ]}\Bigr\}. \nonumber
\end{eqnarray}
First we observe that the definitions differ more when there are inclusions. Indeed, in $P[(\ii)\partial\Lambda]$ the inclusion works also on the derivative; while in $DP[(\ii)\Lambda]$ the derivative works on the inclusions. A consequence of the first observation is that the last term is only a part of the naive variation. The naive variation can be split up like this:
\begin{eqnarray}
STr\Bigl\{ P((\ii)7 \partial \Lambda_6)\Bigr\} &=& STr\Bigl\{\frac12\cdot 7\ D_{a_1}X^{\mu_1}...D_{a_5}X^{\mu_5}[X^{\rho}, X^{\sigma}]\partial_{[\sigma}\Lambda_{\rho \mu_1... \mu_5]}\Bigr\} \label{AJGL-naive1}\\
&=& \frac12STr\Bigl\{ 2\ D_{a_1}X^{\mu_1}...D_{a_5}X^{\mu_5}[X^{\rho}, X^{\sigma}]\partial_{\sigma}\Lambda_{\rho \mu_1... \mu_5}\nonumber \\
& & \quad \quad + 5\ D_{a_1}X^{\mu_1}...D_{a_5}X^{\mu_5}[X^{\rho}, X^{\sigma}]\partial_{\mu_1}\Lambda_{\sigma \rho \mu_2... \mu_5}\Bigr\}.\nonumber
\end{eqnarray}
Inserting (\ref{AJGL-naive1}) into (\ref{AJGL-naive2}) and rearranging the terms yields:
\begin{eqnarray}
\delta \mathcal{L}_{D6}^{trunc2} &=&  STr\Bigl\{ 7 P[\partial \Lambda_6] + 21 i P((\ii)7 \partial \Lambda_6)\label{uitgewerkt}\\
& & + 21i \Bigl( \Lambda_{\mu_1...\mu_6} [F_{[ba_1}X^{\mu_1}]D_{a_2}X^{\mu_2}... D_{a_6]}X^{\mu_6} \nonumber \\
& & \quad \quad + 5\ \Lambda_{\sigma \rho \mu_1 ... \mu_4 }[D_{[b} X^{\rho}, X^{\sigma}]D_{a_1}X^{\mu_1}... D_{a_4}X^{\mu_4}\ F_{a_5 a_6 ]}\nonumber \\
& & \quad \quad -\ \partial_{\sigma}\Lambda_{\rho \eta \mu_1...\mu_4}[X^{\rho}, X^{\sigma}]D_{[b}X^{\eta}D_{a_1}X^{\mu_1}...D_{a_4}X^{\mu_4}\ F_{a_5 a_6]} \Bigr) \nonumber \\
& & -210 ((\ii)\Lambda)_{\mu_1...\mu_4}[F_{[ ba_1}, X^{\mu_1}]D_{a_2}X^{\mu_2}...D_{a_4}X^{\mu_4} F_{a_5 a_6 ]} \Bigr\}\nonumber
\end{eqnarray}
The second, third and fourth line form a single commutator\footnote{See the appendix for comments about commutator manipulations inside the symmetrized trace.}:
\begin{equation}
\begin{array}{l}
\vsp{0.4cm}STr\Bigl\{ [Sym(\Lambda_{\mu_1...\mu_6};F_{[ ba_1};D_{a_2}X^{\mu_2};...;D_{a_6]}X^{\mu_6}),\ X^{\mu_1}] \Bigr\}\\
\vsp{0.2cm}
\quad \quad = STr\Bigl\{ \partial_{\rho}\Lambda_{\mu_1...\mu_6}[X^{\rho}, X^{\mu_1}]F_{[ba_1}D_{a_2}X^{\mu_2}...D_{a_6]}X^{\mu_6}\label{singlecomm1}\\
\vsp{0.2cm}
\quad \quad \quad \quad + \Lambda_{\mu_1...\mu_6}[F_{[ba_1}, X^{\mu_1}]D_{a_2}X^{\mu_2}...D_{a_6]}X^{\mu_6}\nonumber \\
\vsp{0.2cm}
\quad \quad \quad \quad + 5\ \Lambda_{\mu_1...\mu_6}F_{[ba_1}[D_{a_2}X^{\mu_2}, X^{\mu_1}]D_{a_3}X^{\mu_3}...D_{a_6]}X^{\mu_6}\Bigr\}\nonumber.
\end{array}
\end{equation}
But what about the last line of (\ref{uitgewerkt})? It seems that this would form a single commutator together with variations of a term like $P[(\ii)^2C_7]F^2$. Such a term vanishes because there are only three transverse and thus non-Abelian coordinates. If we work out an appropriate single commutator, we get indeed the needed variation and corrections which would fit into a transformation of $P[(\ii)^2C_7]F^2$, but which now just vanish:
\begin{equation}
\begin{array}{l}
\vsp{0.4cm}STr\Bigl\{ [Sym(\Lambda_{\mu_1...\mu_6};[X^{\mu_3}, X^{\mu_2}];F_{[ ba_1};F_{a_2a_3};D_{a_4}X^{\mu_4};...;D_{a_6]}X^{\mu_6}),\ X^{\mu_1}] \Bigr\}\label{singlecomm2}\\
\vsp{0.2cm}
\quad \quad = STr\Bigl\{ \partial_{\rho}\Lambda_{\mu_1...\mu_6}[X^{\rho}, X^{\mu_1}][X^{\mu_3}, X^{\mu_2}] F_{[ba_1}F_{a_2a_3}D_{a_4}X^{\mu_4}...D_{a_6]}X^{\mu_6}\\
\vsp{0.2cm}\quad \quad\quad \quad + \Lambda_{\mu_1...\mu_6}[[X^{\mu_3}, X^{\mu_2}],X^{\mu_1}]F_{[ba_1}F_{a_2a_3}D_{a_4}X^{\mu_4}...D_{a_6]}X^{\mu_6}\\
\vsp{0.2cm}\quad \quad\quad \quad + 2\Lambda_{\mu_1...\mu_6}[X^{\mu_3},X^{\mu_2}][F_{[ba_1}, X^{\mu_1}]F_{a_2a_3}D_{a_4}X^{\mu_4}...D_{a_6]}X^{\mu_6} \\
\vsp{0.4cm}\quad \quad\quad \quad + 3\Lambda_{\mu_1...\mu_6}[X^{\mu_3}, X^{\mu_2}]F_{[ba_1}F_{a_2a_3}[D_{a_4}X^{\mu_4}, X^{\mu_1}]D_{a_5}X^{\mu_5}D_{a_6]}X^{\mu_6}\Bigr\} \\
\quad \quad = STr\Bigl\{ 2\Lambda_{\mu_1...\mu_6}[X^{\mu_3}, X^{\mu_2}][F_{[ba_1}, X^{\mu_1}]F_{a_2a_3}D_{a_4}X^{\mu_4}...D_{a_6]}X^{\mu_6}\Bigr\}.\nonumber
\end{array}
\end{equation}
So the modified variation and the naive one differ by two single commutator terms, which vanish when inside the symmetrized trace.

This is easily generalized to the case of general $C_p$ and numbers of commutators and Born-Infeld field strengths. Thus the other terms appearing in the $D6$ brane action
\begin{equation}
STr \Bigl\{ P[C_{2p+1}]F^{3-p} + p(2p+1)iP[(\ii)C_{2p+1}]F^{4-p} \Bigr\}
\end{equation}
are also invariant under both definitions of gauge transformations.
The only thing yet to do to get the equivalence for the full $D6$ brane action is letting $B$ be arbitrary. This poses no problems at all, since the fields $\mathcal{A}_p$ defined above are invariant under all R-R transformations but the one with parameter $\Lambda_{p-1}$, which is true for both definitions of gauge transformations.
\begin{eqnarray}
\delta_{naive}\ STr\Bigl\{P[(\ii)^r \mathcal{A}_p] \Bigr\}&=& STr \Bigl\{ P[(\ii)^r (p \partial \Lambda_{p-1})] \Bigr\}\\
\delta_{mod}\ STr\Bigl\{P[(\ii)^r \mathcal{A}_p] \Bigr\}&=& STr \Bigl\{ (p-2r) D\Bigl(P[(\ii)^r \Lambda_{p-1}] \Bigr) \Bigr\}\nonumber
\end{eqnarray}
Proving invariance of the general D-brane action under the naive gauge transformation uses the same reasoning as for the simple case of the $D6$.

\sect{The NS-NS variation}\label{sect_NSNS}
The naive and modified R-R variations are equivalent, making the naive definition valid just as in the case of the NS-NS variations. Is the opposite also possible? Namely, can we define an equivalent modified NS-NS variation that looks like the modified R-R transformation? The answer is no. The reason is simple: the NS-NS variation appears multiplied with other fields, unlike the R-R variation. While the difference between the naive and modified definitions of the R-R variation can be arranged into single commutators, the fields multiplying the NS-NS variation makes such an arrangement impossible.
To see this more clearly, we will look again at the $D6$ brane. The NS-NS variation is given by
\begin{eqnarray}
\delta_{naive} P[B]&=& 2P[\partial \Sigma] - 12i P[(\ii)(B\partial \Sigma)] + 2i (\ii)B\ P[\partial \Sigma].\label{naiveB2}
\end{eqnarray}
Only the first term is the proper variation of the field, coming from the Abelian variation $\delta B = 2\partial \Sigma$. A candidate for a modified variation would only differ in that first term. The other two, being variations of the worldvolume fields $V$ and $X$, remain as they are.
So the candidate modified transformation looks like
\begin{eqnarray}
\delta_{mod} P[B]&=& 2DP[\Sigma] - 12i P[(\ii)(B \partial \Sigma)] + 2i (\ii)B \ P[\partial \Sigma].\label{ajglB}
\end{eqnarray}
To avoid writing more terms than necessary, we will look just at the difference between the two definitions. Only the variation of $B$ itself changes, and the differences are:
\begin{eqnarray}
(\delta_{mod}-\delta_{naive})P[B]&=& i[F,X^{\mu}]\Sigma_{\mu}\\
(\delta_{mod}-\delta_{naive})P[[X^{\rho}, X^{\sigma}]C_{\sigma ...}B_{\rho}]&=& P[[DX^{\rho}, X^{\sigma}]C_{\sigma ...}\Sigma_{\rho}]\nonumber\\
(\delta_{mod}-\delta_{naive})(\ii)B &=& -(\ii)\partial \Sigma.\nonumber
\end{eqnarray}
It was proven that under the variation (\ref{naiveB2}), blocks with the same R-R field are invariant, like the $D6$ block
\begin{eqnarray}
\mathcal{L} &=& STr \Bigl\{ 21P[C_5B]+21P[C_5]F\\
& & + 378i P[(\ii)(C_5 B^2)] + 411i P[(\ii)(C_5 B)]F + 105iP[(\ii)C_5]F^2 \Bigr\}.\nonumber
\end{eqnarray} 
We will now apply the candidate modified definition to this block.
\begin{eqnarray}
\delta_{mod} \mathcal{L} &=& (\delta_{mod}-\delta_{naive})\mathcal{L}+\delta_{naive}\mathcal{L}\\
&=& STr \Bigl\{ 21i D_{[a_1}X^{\mu_1}...D_{a_5}X^{\mu_5}[F_{a_6 a_7]}, X^{\rho}]\Sigma_{\rho}C_{\mu_1...\mu_5}\nonumber\\
& & -21 i D_{[a_1}X^{\mu_1}...D_{a_5}X^{\mu_5} F_{a_6a_7]}[X^{\rho},X^{\sigma}]\partial_{\sigma}\Sigma_{\rho}C_{\mu_1 ... \mu_5}\nonumber\\
& & +210 i D_{[a_1}X^{\mu_1}...D_{a_4}X^{\mu_4}F_{a_5a_6}[D_{a_7]}X^{\rho}, X^{\sigma}]C_{\sigma \mu_1...\mu_4}\Sigma_{\rho}\nonumber\\
& & + \textrm{terms proportional to $B$ or $F^2$}\Bigr\}.\nonumber
\end{eqnarray}
The above three terms do not form a single commutator, already because there will never be a variation term proportional to $[C_5,X]$.
So, while the $D6$ brane action is invariant under NS-NS variations defined naively for the $B$ form, along with the variations of the worldvolume fields, invariance is impossible for our candidate modified transformation. 

\sect{Discussion}\label{sect_disc}
In this paper it is proven that the modified definition of the R-R gauge variations gives the same result as the naive one up to a single commutator. While invariance is more manifest when using the modified definition, the naive variation is easier when regarding dualities. In particular, we see that the S-dual twoforms $C_2$ and $B$ have the same gauge properties.

Equivalence is not the case for the NS-NS transformations. A candidate modified transformation can be thought of, but the difference with the naive transformation can not be arranged into a single commutator or anything else vanishing. This means that the multiple D-brane actions are not invariant under modified NS-NS variations.

One can take the naive definition as definition of gauge transformations for both $C$ and $B$ fields and use the modified R-R transformation to prove the invariance of the multiple brane's Chern-Simons action. 

\vsp{1cm}

\noindent
{\bf Acknowledgments}\\
\vspace{-.2cm}

\noindent
We wish to thank Walter Troost and especially Bert Janssen for the useful discussions.
This work is supported in part by the European Community's Human
Potential Programme under contract MRTN-CT-2004-005104 `Constituents,
fundamental forces and symmetries of the universe'. J.A. is Aspirant FWO Vlaanderen.
Her work is supported in part by the FWO - Vlaanderen, project
G.0235.05 and by the Federal Office for Scientific, Technical and
Cultural Affairs through the "Interuniversity Attraction Poles Programme
-- Belgian Science Policy" P5/27.

\appendix
\renewcommand{\theequation}{\Alph{section}.\arabic{equation}}

\sect{Symmetrized trace calculations}
The symmetrized trace prescription consists of symmetrizing all entries followed by taking the trace. Hereby commutators are seen as one entry.
So, if $A,B,C,D$ and $E$ are matrices,
\begin{eqnarray}
STr\Bigl\{ ABCDE\Bigr\}&=& Tr\Bigl\{ Sym\Bigl(A; B; C;D;E \Bigr)\Bigr\}\\
STr\Bigl\{[A,B]CD\Bigr\}&=& Tr\Bigl\{ Sym\Bigl( [A,B];C;D \Bigr)\Bigr\}\nonumber\\
STr\Bigl\{ [AB,C]DE \Bigr\}&=& Tr \Bigl\{ Sym\Bigl( [AB,C];D;E\Bigr)\Bigr\}.\nonumber
\end{eqnarray}
Symmetrizing is denoted by $Sym$:
\begin{eqnarray}
Sym\Bigl(A; B; C ;D;E\Bigr)&=& ABCDE + ABCED + \textrm{other permutations},
\end{eqnarray}
Due to the behavior of commutators within the symmetrized trace, care is needed when using common commutator manipulations. Indeed, simply substituting $AB-BA$ for $[A,B]$ is already problematic. While
\begin{eqnarray}
STr\Bigl\{[A,B]CD\Bigr\}&=& Tr\Bigl\{ Sym\Bigl( [A,B];C;D \Bigr)\Bigr\}
\end{eqnarray}
is in general nonzero, the substitution would make it vanish identically:
\begin{eqnarray}
STr\Bigl\{ (AB - BA)CD\Bigr\}&=& Tr\Bigl\{ Sym\Bigl( A;B;C;D \Bigr) - Sym\Bigl(B;A;C;D \Bigr)\Bigr\} = 0.
\end{eqnarray}

What about splitting the commutator of a product, $[AB,C] = A[B,C]+[A,C]B$? One can expect that the rule does not hold when it is multiplied by other matrices. Indeed, on the left side the $A,B$ and $C$ will stay together, while on the right $A$ and $B$ will permute among the other matrices:
\begin{eqnarray}
STr\Bigl\{[AB,C]DE\Bigr\} &=& Tr \Bigl\{ Sym\Bigl( [AB,C];D;E\Bigr)\Bigr\}\\
&=& Tr \Bigl\{ Sym\Bigl( A[B,C] + [A,C]B;D;E\Bigr)\Bigr\}\nonumber 
\end{eqnarray}
and
\begin{eqnarray}
STr\Bigl\{ A[B,C]DE + [A,C]BDE\Bigr\} &=& Tr \Bigl\{ Sym\Bigl( A;[B,C];D;E\Bigr) + Sym\Bigl( [A,C];B;D;E\Bigr) \Bigr\}.
\end{eqnarray}
 
The problem only appears when there are at least two other matrices to permute with. In the case of a single commutator, or with one extra factor, the symmetrized trace reduces to an ordinary trace and the rule is valid. The product inside the commutator needs to be symmetrized itself, though.
\begin{eqnarray}
STr\Bigl\{ [Sym(A;B;C),D]E\Bigr\} &=& Tr\Bigl\{ [Sym(A;B;C),D]E \Bigr\}\\
&=& Tr \Bigl\{ [A,D]Sym(B;C;E)+ [B,D]Sym(A;C;E)+ [C,D]Sym(A;B;E)\Bigr\}\nonumber \\
&=& STr \Bigl\{ [A,D]BCE + A[B,D]CE + AB[C,D]E\Bigr\}.\nonumber
\end{eqnarray}
In going from the second to the third line, the cyclic property of the trace has been used.


\begin{thebibliography}{99}
\bibitem{Witten} E. Witten, {\it Bound States Of Strings And $p$-Branes}, Nucl. Phys. B460 (1996) 335, hep-th/9510135.

\bibitem{Douglas} M. Douglas, {\it Branes within Branes}, hep-th/9512077.

\bibitem{GM} M. Garousi, R. Myers, {\it World-Volume Interactions on D-Branes}, Nucl. Phys. B542 (1999) 73, hep-th/9809100.

\bibitem{Dorn} H. Dorn, {\it Nonabelian gauge field dynamics on matrix D-branes}, Nucl. Phys. B494 (1997) 105, hep-th/9612120.

\bibitem{Hull} C. Hull, {\it Matrix Theory, U-Duality and Toroidal Compactifications of M-Theory}, JHEP 9810 (1998) 011, hep-th/9711179.

\bibitem{Tseytlin} A. Tseytlin, {\it On non-Abelian generalization of Born-Infeld action in string theory}, Nucl. Phys. B501 (1997) 41, hep-th/9701125.

\bibitem{Myers} R. Myers, {\it Dielectric-Branes}, JHEP 9912 (1999) 022, hep-th/9910053.

\bibitem{TvR1} W. Taylor, M. Van Raamsdonk, {\it Multiple D0-branes in weakly curved backgrounds}, Nucl. Phys. B558 (1999) 26, hep-th/9904095.

\bibitem{TvR2} W. Taylor, M. Van Raamsdonk, {\it Multiple Dp-branes in weak background fields}, Nucl. Phys. B573 (2000) 703, hep-th/9910052.

\bibitem{TV} S. Trivedi, S. Vaidya, {\it Fuzzy Cosets and their Gravity Duals}, JHEP 0009 (2000) 041, hep-th/0007011.

\bibitem{vR} M. Van Raamsdonk, {\it Blending loca symmetries with matrix nonlocality in D-brane effective actions}, JHEP 0309 (2003) 026, hep-th/0305145.

\bibitem{AJGL} J. Adam, J. Gheerardyn, B. Janssen, Y. Lozano, {\it The gauge invariance of non-Abelian Chern-Simons action for D-branes revisited}, Phys.Lett B589 (2004) 59-69, hep-th/0312264.

\bibitem{dBS} J. de Boer, K. Schalm, {\it General covariance of the non-abelian DBI-action}, JHEP 0302 (2003) 041, hep-th/0108161.

\bibitem{BFLvR} D. Brecher, K. Furuuchi, H. Ling, M. Van Raamsdonk, {\it Generally Covariant Actions for Multiple D-Branes}, JHEP 0406 (2004) 020, hep-th/0403289.

\bibitem{BKLvR} D. Brecher, P. Koerber, H. Ling, M. Van Raamsdonk, {\it Poincar\'e invariance in Multiple D-brane Actions}, hep-th/0509026.

\bibitem{AIJ} J. Adam, I.A.Ill'an, B. Janssen, {\it On the gauge invariance and coordinate transformations of non-Abelian D-brane actions},JHEP10 (2005) 022, hep-th/0507198.

\end{thebibliography}
\end{document}